\documentstyle[prl,twocolumn,aps,epsf]{revtex}

\begin{document}

\hoffset-1cm

\draft
\preprint{nucl-th/9704???}

\title{Bose-Einstein Weights for Event Generators}

\author{Q.H. Zhang\cite{home}, U.A. Wiedemann, C. Slotta, and U. Heinz}

\address{
   Institut f\"ur Theoretische Physik, Universit\"at Regensburg,\\
   D-93040 Regensburg, Germany
}

\date{\today}

\maketitle

\begin{abstract}
A simple new algorithm for the calculation of two-particle Bose-Einstein
correlations from classical event generators is derived and discussed.
\end{abstract} 


%%%%%%%%%%%%%%%%%%%%%%%%%%%%%%%%%%%%%%%%%%%%%%%%%%%%%%%%%%%%%%%%%%%%%%
% Beginning of text
%%%%%%%%%%%%%%%%%%%%%%%%%%%%%%%%%%%%%%%%%%%%%%%%%%%%%%%%%%%%%%%%%%%%%%

For ultrarelativistic heavy-ion collisions, two-particle Bose-Einstein 
correlations between identical pions or kaons provide a unique 
possibility to reconstruct the geometry (size, temporal extension)
${\em and}$ dynamics (collective expansion flow) of the source at the 
point of hadron freeze-out. (For a recent theoretical review see 
\cite{H96}.) This reconstruction is, however, not completely model 
independent; it requires the use of ``reasonable'' source 
parametrizations (e.g. \cite{H96,CL96,CN96}) whose parameters are 
then fixed by a simultaneous analysis of single-particle momentum 
spectra and two-particle momentum correlations \cite{CN96,S96}.

Invaluable help for the selection of ``reasonable'' source 
parametrization comes from microscopic event generators (e.g. VENUS 
\cite{W93}, RQMD \cite{SSG89} or ARC \cite{PSK92}) which generate the 
phase-space distribution of hadrons at freeze-out from a dynamical 
Monte Carlo simulation of the (classical) kinetic phase-space 
evolution of the collision zone. Unfortunately, it was recently 
pointed out \cite{A97,MKF96} that the direct computation of two-particle 
correlation functions from classical kinetic codes \cite{W93,SSG89,PSK92} 
is fraught with severe conceptual and practical problems. These can be 
simply understood by starting from the general relation \cite{S73}
between the 2-particle correlation function $C(\bbox{q}, \bbox{K})$ 
and the (real) ``emission function'' (one-particle Wigner density at 
freeze-out) $S(x,K)$, 
 \begin{mathletters}
 \label{1} 
 \begin{eqnarray}
 \label{1a} 
   \FL C(\bbox{q},\bbox{K}) =
   1 + \frac{{\left\vert\int d^{4}x\, S(x,K)\, e^{iq\cdot x}
              \right\vert}^{2}} 
            {\int d^{4}x\, S(x,p_{a}) \,\int d^{4}y\, S(y,p_{b})} \, ,
 \\
 \label{1b}
   \FL \bbox{q}{=}\bbox{p}_a{-}\bbox{p}_b, \ 
       q^{0}{=}E_{a}{-}E_{b},\ 
       \bbox{K}{=}{\bbox{p}_a{+}\bbox{p}_b \over 2},\
       K^{0}{=}{E_{a}{+}E_{b} \over 2}.
 \end{eqnarray}
 \end{mathletters}
\noindent
$(\bbox{p}_a, E_{a})$ and $(\bbox{p}_b, E_{b})$ are the 4-momenta of 
the two observed particles. Eq.~(\ref{1a}) neglects final 
state interactions which we will leave out in this Letter in order to 
concentrate on the principal issues.  

The numerator in the second term of Eq.~(\ref{1a}) can be rewritten as
 \begin{eqnarray}
 \label{2}
   {\rm Num} (q,K) =&& 
   \int d^4x\, d^4y \,S(x,K)\, S(y,K) 
 \nonumber\\
   && \ \times\, \cos(q{\cdot}(x-y)) \, .
 \end{eqnarray}
The problem is the construction of the Wigner density $S(x,K)$ from 
the output of the event generators. The latter consists of a set of 
phase-space points $(x_{i},p_{i})$ denoting the (on-shell) momenta
$p_{i}$ and points of last interaction $x_{i}$ of the produced particles. 
According to (\ref{1b}) $K$ is the average of two on-shell momenta, but 
not itself on-shell, $K^{0} \ne \sqrt{m^2+{\bbox{K}^2}}$. Therefore 
$S(x,K)$ cannot be directly related to the phase-space density 
generated by the distribution of points $(x_{i},p_{i})$. To overcome 
this difficulty one ususally imposes \cite{YK78,P94} the ``smoothness 
assumption'' $S(x,\frac{p_a+p_b}{2}) \approx S(x,p_a) \approx S(x,p_b)$ 
and rewrites the expression 
(\ref{2}) as
 \begin{eqnarray}
 \label{3}
   {\rm Num}(q,K) =&& 
   \int d^{4}x\, d^{4}y \, S(x,p_a) \, S(y,p_b) 
 \nonumber\\   
    && \times \cos\bigl((p_a-p_b){\cdot}(x-y)\bigr)\, .
 \end{eqnarray}
One now identifies $S(x,p)$ with the classical output distribution 
from the event generator,
 \begin{equation}
 \label{4}
   S_{\rm class}(x,p) = \sum_{i=1}^{N} \delta^{(4)}(x-x_{i})\, 
   \delta^{(4)}(p-p_{i})\, ,
 \end{equation}
where $N$ is the total number of pions of a given charge in the 
event. In the widely used Pratt algorithm \cite{P94} the expression 
which results after inserting (\ref{4}) into (\ref{3}) is simulated 
by the ad hoc prescription
 \begin{equation}
 \label{5}
   {\rm Num} (q,K) \longmapsto \sum_{i,j \in {\rm bin}} 
   \cos\bigl((p_i-p_j){\cdot}(x_i-x_j)\bigr) \, .
 \end{equation}
Here ``bin'' denotes a small bin in $(\bbox{q},\bbox{K})$ with 
$\bbox{p}_i-\bbox{p}_j \approx \bbox{q}$ and $(\bbox{p}_i+\bbox{p}_j)/2
\approx \bbox{K}$. (In practice the bin size depends on event 
statistics.)

The prescription (\ref{5}) has two severe problems: first, the positivity 
of Num$(q,K)$ got lost between Eqs.~(\ref{2}) and (\ref{3}) when making 
the ``smoothness assumption''. It was pointed out in Refs.~\cite{CH94,PRW94}
and practically demonstrated in Ref.~\cite{MKF96} that for sources with 
strong $x$-$p$-correlations (e.g. rapidly expanding sources) this can 
lead to unphysical oscillations of the simulated correlation function 
around unity. Second, the intuitive substitution law 
(\ref{5}) is formally incorrect and results in a wrong selection of 
contributing pairs as well as an incorrect weighting factor for each 
pair. 

To prove this last point let us write in (\ref{2})
 \begin{eqnarray}
 \label{6}
   S(x,K)\!\!&&\!\!\! S(y,K) = \int d^4P_1\, d^4P_2 \, S(x,P_1)\, S(y,P_2) 
 \nonumber\\
   &\times&\, \delta^{(4)}(P_1-P_2) \,
              \delta^{(4)}\left(K-\frac{P_1+P_2}{2}\right)\, .
 \end{eqnarray} 
Inserting the classical expression (\ref{4}) one finds instead of (\ref{5})
 \begin{mathletters}
 \label{7}
 \begin{eqnarray}
 \label{7a}
   {\rm Num}(q,K) =&& \sum_{i,j=1}^{N} \delta^{(4)}(p_i-p_j) \,
   \delta^{(4)}\left(K-\frac{p_i+p_j}{2}\right) 
 \nonumber\\
   && \qquad \times \cos\bigl(q{\cdot}(x_i-x_j)\bigr) 
 \\
 \label{7b}
   \longmapsto && 
   \sum_{i,j \in {\rm bin}(K,\epsilon)} \cos\bigl(q{\cdot}(x_i-x_j)\bigr)\, .
 \end{eqnarray}
 \end{mathletters}
Here ``${\rm bin}(K,\epsilon)$'' denotes a small bin around $K$ with 
width $\epsilon$ in each of the four directions. The prescription 
(\ref{7b}) can be derived rigorously and directly by first generating
from Eq.~(\ref{4}) a piecewise constant function (``histogram'') through
``binning'',
 \begin{eqnarray}
 \label{binning}
   \bar S (x,K) &=& \int_{{\rm bin}(K,\epsilon)} d^4p\, S_{\rm class}(x,p)
 \nonumber\\
   &=& \int_ {K-\epsilon/2}^{K+\epsilon/2} d^4p\, S_{\rm class}(x,p)\, ,
 \end{eqnarray}
and then inserting $\bar S(x,K)$ into Eq.~(\ref{6}). Eq.~(\ref{binning})
is a technical step required by finite event statistics; in practice,
$\epsilon$ should be chosen as small as technically possible. Note that the 
selection of pairs in (\ref{7b}) differs from the one in (\ref{5}):
for given $K$ the algorithm (\ref{7}) selects pairs with $p_i \approx 
p_j \approx K$, independent of the value $\bbox{q}$ at which the 
correlation is to be evaluated. For different values of $\bbox{q}$ 
at fixed $\bbox{K}$, the correlator is obtained by weighting the 
{\em same} set of pairs with different weight factors 
$\cos\bigl(q{\cdot}(x_i-x_j)\bigr)$ which depend only on the spatial
coordinates, but not on the momenta of the particles in the pair. This 
is consistent with the expectation from Eqs.~(\ref{1a},\ref{2}) that 
the measured $\bbox{q}$-dependence of the correlator gives access to 
the distribution of relative distances $x_i-x_j$ in the source (at 
fixed $K$). Since the steps from Eq.~(\ref{2}) to Eq.~(\ref{7}) involve 
only identical transformations (the difference between (\ref{7a}) and 
(\ref{7b}) arising only from a different choice of emission functions 
(\ref{4}) resp. (\ref{binning})), they preserve the positivity of the 
second term in (\ref{1a}). Please note that in contrast to (\ref{5}), 
(\ref{7}) is a continuous function of $\bbox{q}$, i.e. no binning in 
$\bbox{q}$ is required.  

We have checked for the simple analytically solvable model presented 
in \cite{MKF96} that the algorithm (\ref{7}) indeed allows to reconstruct
from a classical Monte Carlo simulation the correct analytical expression 
for $C(\bbox{q},\bbox{K})$; in particular it removes the unphysical 
oscillations found in \cite{MKF96}. In Fig.~1 we show various 
approximations for the correlation function for a 1-dimensional source 
in $z$-direction with emission function 
 \begin{equation}
 \label{model}
  S(z,t;K) = e^{-z^2/R^2}\, \delta(K-\alpha z)\, 
                                   \delta(t)\, ,
 \end{equation}
with $R{=}10$ fm and $\alpha{=}0.02$ GeV/fm. This classical source 
features perfect $z$-$K$-correlations, $K{=}\alpha z$, which are, of 
course, quantum mechanically forbidden (see below). For the source 
(\ref{model}), the exact correlator (\ref{1}) can be calculated 
analytically, yielding $C(q) = 1 + \exp[q^2/(2\alpha^2 R^2)]$ 
(solid line in Fig.~1). The pathological rise of $C(q)$ {\em 
above} the value 2 at $q{=}0$ is due to the violation of 
the uncertainty relation between $z$ and $p_z$ by the model 
(\ref{model}); we selected this model because we believed that such a 
feature may be particularly difficult to reproduce in an event 
generator. Indeed, reconstructing the correlator from a Monte Carlo 
simulation of (\ref{model}) via the Pratt prescription (\ref{5}) 
yields the long-dashed line in Fig.~1 \cite{MKF96}; it can be 
calculated analytically from Eq.~(\ref{3}) as $C(q) = 1 + 
\cos(q^2/\alpha)$. This is always less than 2, but oscillates wildly 
around 1, becoming even 0 at regular $q$-intervals. This contradicts 
the positivity of the second term in (\ref{1}) and is due to the 
smoothness approximation (\ref{3}). -- The two remaining lines in 
Fig.~1 show results from the same Monte Carlo simulation of 
(\ref{model}) but reconstructing the correlator through the new 
algorithm (\ref{7b}) instead of (\ref{5}). For the dot-dashed line 
the bin width $\epsilon$ was chosen as $\epsilon= 10$ MeV, for the 
dotted line as $\epsilon = 5$ MeV. In both cases the simulated result 
deviates from the exact one (solid line); this is not a failure 
of the algorithm, but a result of the binning procedure (\ref{binning}) 
applied to the source (\ref{model}) -- a purely technical step required 
by finite event statistics. As seen, the discrepancy decreases with 
decreasing bin width $\epsilon$.

\vskip -4.0cm
%%%%%%%%%%%%%%%%%%%%%%%%%%%%%%%%%%%%%%%%%%%%%%%%%%%%%%%%%%%%%%%%%%%%
\begin{figure}[h]\epsfxsize=10cm 
\centerline{\epsfbox{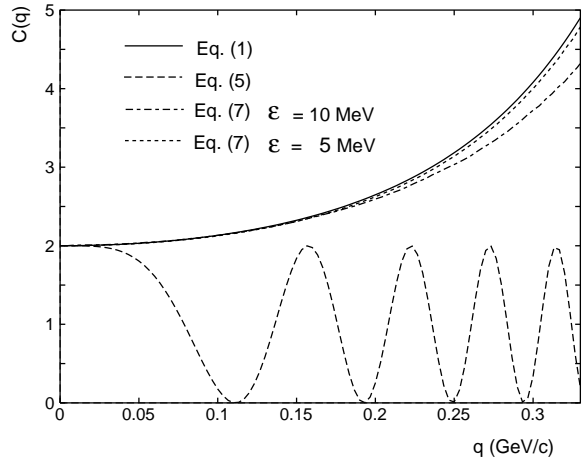}}
\vskip -2.5cm
\caption{\it 
Two-particle correlation function for the model source 
(\protect\ref{model}). Different curves show different
approximations as described in the text. The statistical errors of
the simulation are below the line widths.
}\label{fig1}
\end{figure}
%%%%%%%%%%%%%%%%%%%%%%%%%%%%%%%%%%%%%%%%%%%%%%%%%%%%%%%%%%%%%%%%%%%

While Eq.~(\ref{7}) thus solves the technical problems of the 
prescription (\ref{5}), it does not address the principal physical 
problem that the classical distribution (\ref{4}) is not a valid 
Wigner density since it violates the uncertainty relation by 
simultaneously fixing the coordinates $x_{i}$ and momenta $p_{i}$ of 
the emitted particles. It has been repeatedly suggested 
\cite{PGG90,Z96,MP97} that this can be remedied by replacing the 
sharply localized $\delta$-functions in (\ref{4}) by 
minimum-uncertainty wave packets. In the remainder of this Letter we 
will discuss how such a procedure will modify the algorithm for 
calculating single-particle spectra and two-particle correlations from 
event generators, thereby rendering it quantum mechanically 
consistent.  

Let us start from the folding relation for the emission function, 
derived in Ref.~\cite{CH94} within the covariant current formalism 
\cite{GKW79}:
 \begin{equation}
 \label{8}
   S(x,K) = \int d^4z \, d^4Q \, \rho(x-z,Q) \, S_{0}(z,K-Q) \, .
 \end{equation}
Here 
 \begin{equation}
 \label{9}
   S_{0}(x,p) = \int d^4v\, e^{-ip{\cdot}v}\, 
   j_0^*\left(x+{\textstyle{v\over 2}}\right) \, 
   j_0\left(x-{\textstyle{v\over 2}}\right)
 \end{equation}
is the Wigner density associated with an elementary source current 
amplitude $j_0(x)$, taken below as a Gaussian wavepacket, and 
$\rho(x,p)$ is a classical phase-space distribution for the 
centers $x_i$ and (average) momenta $p_i$ of these wave packets, here 
taken as
 \begin{equation}
 \label{10}
   \rho(x,p) = \sum_{i=1}^N \delta^{(4)}(x-x_i) \, \delta^{(4)}(p-p_i)\, .
 \end{equation}
For the elementary source amplitude we make the ansatz
 \begin{equation}
 \label{11}
   j_0(x) = {\cal N} \, \exp\left(-{\bbox{x}^2 \over 2\sigma^2}\right)
   \, \delta(x^0) \, ;
 \end{equation}
this source emits a Gaussian wave-packet with width parameter $\sigma$
at ``freeze-out time'' $x^0$. With this ansatz the elementary Wigner density 
$S_0$ becomes
 \begin{equation}
 \label{12}
   S_0(x,p) =  8 (\pi\sigma^2)^{\frac{3}{2}} \,\vert{\cal N} \vert^2 
   \delta(x^{0}) \exp\left(-{\bbox{x}^2 \over \sigma^2}
                              -\sigma^2\bbox{p}^2 \right) \, .
 \end{equation}
Inserting this and (\ref{10}) into (\ref{8}) one finds
 \begin{eqnarray}
 \label{13}
   S(x,K) = {\cal N'} \sum_{i=1}^N &&\delta(x^0-x_i^0) \,
   \exp\left(-{(\bbox{x}-\bbox{x}_i)^2 \over \sigma^2}\right)
 \nonumber\\
   &\times& 
     \exp\left(-\sigma^2(\bbox{K}-\bbox{p}_i)^2 \right)\, , 
\end{eqnarray}
with ${\cal N'} = 8(\pi\sigma^2)^{\frac{3}{2}} \vert {\cal N}\vert^2$.
This generalizes the classical ansatz (\ref{4}) into a quantum mechanically 
consistent source Wigner density; with the free parameter $\sigma$ one
can choose the relative degree of localization in coordinate space 
$(\sigma \to 0)$ or momentum space $(\sigma \to \infty)$, always 
preserving $\Delta x \cdot \Delta p = \hbar$.

The single particle spectrum, which occurs in the denominator of the 
second term of (\ref{1a}), is now given by 
 \begin{mathletters}
 \label{14}
 \begin{eqnarray}
 \label{14a}
   E_a {dN\over d^3p_a} &=& \int d^4x \, S(x,p_a)  = 
   {\cal N''} \sum_{i=1}^{N} v_i(\bbox{p}_a) \, ,
 \\
 \label{14b}
   {\cal N''} &=& (2\pi\sigma^2)^2 \vert {\cal N} \vert^2 \, ,
 \\
 \label{14c}
   v_ i (\bbox{p}_a) &=& 
   \exp\bigl(-\sigma^2 (\bbox{p}_a-\bbox{p}_i)^2\bigr) \, ,
 \end{eqnarray}
 \end{mathletters}
\noindent
and similarly for $\bbox{p}_b$. It is normalized to the total number $N$ 
of pions in the event; this fixes the normalization constant ${\cal N}$ 
above. The exchange term (\ref{2}) in the two-particle spectrum (with 
$q$ and $K$ defined in Eq.~(\ref{1b})) is similarly derived as
 \begin{mathletters}
 \label{15}
 \begin{eqnarray}
 \label{15a}
   {\rm Num}(\bbox{q},\bbox{K})\! &=& \!({\cal N''})^2 
   \exp(-{\textstyle{1\over 2}}\sigma^2\bbox{q}^2) 
   \sum_{i,j=1}^N w_{ij}(\bbox{q},\bbox{K})  ,
 \\
 \label{15b}
   w_{ij}(\bbox{q},\bbox{K})\! &=& \! v_i(\bbox{K}) \, v_j(\bbox{K}) \, 
   \cos\bigl(q{\cdot}(x_i-x_j)\bigr)  .
 \end{eqnarray}
 \end{mathletters}
\noindent
The normalization ${\cal N''}$ drops out in the correlator (\ref{1a}). 
Eq.~(\ref{15}) should be compared with the classical expressions (\ref{5}) 
and (\ref{7}). Like (\ref{7}) (but contrary to (\ref{5})) it is 
positive definite and thus free of spurious oscillations around 0. 
The sum in (\ref{15a}) is now over {\em all} pairs $(i,j)$; the sharp 
restriction to the bin ``${\rm bin}(K,\epsilon)$'' in (\ref{7}) is replaced 
by the Gaussian weight factors $v_i(\bbox{K})$ and $\exp(-{1\over 2} 
\sigma^2 \bbox{q}^2)$. Please note, however, that the former also 
occur in the new definition (\ref{14a}) for the single particle 
spectrum, and must be kept in both places for consistency.  
Eq.~(\ref{15b}) shares with Eq.~(\ref{7}) the property (which was 
already discussed) that the cosine weight factor for each pair $(i,j)$ 
depends only on $x_i-x_j$, but not on $p_i$ and $p_j$. By combining
Eqs.~(\ref{14}) and (\ref{15}) it can be shown that now the correlator
$C(\bbox{q,K})$ is always between 1 and 2, i.e. that the pathological
rise above 2 shown in Fig.~1 cannot happen for an emission function 
which respects the uncertainty relation. In Ref.~\cite{Wetal} the 
results (\ref{14}) and (\ref{15}) (but not (\ref{7})) were derived with 
different methods, including finite multiplicity corrections.  

Expressions (\ref{14}) and (\ref{15}) depend on one free parameter, 
the Gaussian width $\sigma$ of the wave-packets. It is instructive
to discuss the two obvious limits, $\sigma \to 0$ and $\sigma \to 
\infty$. For $\sigma=0$ the elementary sources are sharply localized in 
space (cf. Eq.~(\ref{11})); as a result, the single-particle momentum 
spectrum (\ref{14}) is completely flat. Furthermore, one sees from
Eqs.~(\ref{15}) that the correlation function becomes $K$-independent, 
even for expanding sources of type (\ref{10}) where the $x_i$ and $p_i$ 
are strongly correlated. Both features are clearly unrealistic. In 
the opposite limit, $\sigma \to \infty$, the Gaussian smearing factors 
in the single particle spectrum (\ref{14}) disappear, and Eq.~(\ref{14a})
degenerates to a sum over $\delta$-functions; this is the usually 
employed algorithm for computing single-particle spectra from event 
generators with classical particle propagation \cite{W93,SSG89,PSK92}. 
The two-particle correlation term (\ref{15}), on the other hand, is then
sharply concentrated at $q=0$, i.e. the correlator $C(\bbox{q},\bbox{K})$ 
drops from 2 to 1 over a $q$-range of order $1/\sigma$. This translates 
into a source radius $\sim \sigma$ and reflects the diverging spatial 
extension of the elementary wavepackets in this limit, irrespective of 
the (localized) spatial distribution $\rho(x,p)$ of their centers.

It is thus clear that in practice $\sigma$ must be kept finite, but 
non-zero. As pointed out in \cite{PGG90,Z96,MP97} this implies a 
broadening of the single-particle momentum distributions relative to 
the one derived from the classical phase-space distribution $\rho (x,p)$ 
of freeze-out points. Using the algorithms (\ref{14},\ref{15}) for a 
quantum mechanically consistent computation of spectra and Bose-Einstein 
correlations from classical event generators thus requires a retuning 
of the codes to elementary $e^+e^-$ and $pp$ collisions, using the same 
algorithms there.

This last step imposes rather restrictive limits for the value of 
$\sigma$ \cite{PGG90}. Since the pions from high energy $pp$ and 
$e^+e^-$ collisions have an average transverse momentum $\langle 
p_\perp \rangle \simeq 0.35$ GeV, the Gaussian width in momentum space 
of the elementary wave packets must be below this value. This implies 
$\sigma \gtrsim 0.5$ fm. On the other hand, the effective source 
radii for such collisions extracted from 2-particle correlations are 
of the order of only $0.8 - 2$ fm \cite{L89}. This implies $\sigma 
\lesssim 1$ fm. It was already pointed out in the pioneering GGLP 
paper \cite{GGLP60} that this value roughly agrees with the pion's Compton 
wavelength. This suggests the following interpretation of the measured 
pion spectra and Bose-Einstein correlations from elementary hadron-hadron 
and $e^+e^-$ collisions: each elementary collision produces a small 
number of elementary Gaussian wave packets with width $\sigma$, 0.5 fm 
$< \sigma <$ 1 fm. 
%
%For the simplest case of just one wavepacket
%with c.m. momentum $\bbox{p}_1$ the single particle spectrum and 
%two-particle correlation function read
% \begin{mathletters}
% \label{16}
% \begin{eqnarray}
% \label{16a}
%   E {dN\over d^3p} &=& {\cal N}'' 
%      \exp\bigl( - \sigma^2 (\bbox{p}-\bbox{p}_1)^2 \Bigr) \, ,
% \\
% \label{16b}
%   C(\bbox{q},\bbox{K}) &=& 1 + \exp(-{\textstyle{1\over 2}}\sigma^2 
%   \bbox{q}^2) \, .
% \end{eqnarray}
% \end{mathletters}
For $\sigma \sim 0.5$ fm the width of the Gaussian momentum 
distribution in (\ref{14c}) nearly exhausts the measured $\langle p_\perp 
\rangle$; the measured single-pion spectra thus reflect mostly the 
intrinsic momentum distribution of the elementary pion wave packet.  
On the other hand, the HBT radius $R_{\perp}^{\rm HBT} = 
\sigma/\sqrt{2}$ corresponding to the $q$-Gaussian in (\ref{15a}) 
nearly exhausts the values extracted from two-particle correlation 
measurements; the latter thus mirror the intrinsic spatial width of 
the wave packets. There is very little additional room in the data for 
random (thermal) motion of the wave packets relative to each other, 
nor for their spatial distribution over a larger volume. If the total 
source is much bigger than 1 fm, it must expand very rapidly, with 
homogeneity regions which are not much larger than the size of an 
elementary wave packet.  In this sense, pion spectra from $e^{+}e^{-}$ 
and $pp$ collision measure the smallest sources compatible with the 
uncertainty relation.  

In summary, we have derived a new algorithm for the computation of 
single-particle spectra and two-particle correlations from classical event 
generators. The classical algorithm (\ref{7}) removes the recently 
discovered deficiencies of the presently employed Pratt algorithm (\ref{5}).
The quantum mechanical algorithm (\ref{14},\ref{15}) additionally 
ensures that the uncertainty relation is not violated. We also showed 
how the free parameter $\sigma$ in the latter algorithm can be fixed 
from elementary $e^{+}e^{-}$ and hadron-hadron collisions. 

We acknowledge stimulating conversations with P. Foka, H. Kalechofsky,
M. Martin, and S. Pratt, as well as many discussions during the HBT96 
Workshop at the ECT* in Trento, Sept. 16-27, 1996, which brought the 
problems with (\ref{5}) into the open. This work was supported by BMBF, 
GSI and DFG. Q.H.Z. acknowledges support by the Alexander von Humboldt 
Foundation through a Research Fellowship.

%%%%%%%%%%%%%%%%%%%%%%%%%%%%%%%%%%%%%%%%%%%%%%%%%%%%%%%%%%%%%%%%%%%%%%


\begin{thebibliography}{99}
\bibitem[*]{home} 
  Humboldt Research Fellow; on leave from China Center of
  Advanced Science and Technology (CCAST). 
\bibitem{H96} 
  U. Heinz, in: {\it Correlations and Clustering Phenomena in 
  Subatomic Physics}, ed. by M.N. Harakeh, O. Scholtan and J.H. Koch, 
  NATO ASI Series B, (Plenum, New York, 1997), in press (Los Alamos 
  eprint archive nucl-th/9609029)
\bibitem{CL96} 
  T. Cs\"org\H o and B. L\"orstad, Phys. Rev. C{\bf 54} (1996) 1396;
  and Nucl. Phys. A{\bf 590} (1995) 465c.
\bibitem{CN96} 
  S. Chapman and J.R. Nix, Phys. Rev. C{\bf 54} (1996) 866.
\bibitem{S96} 
  S. Sch\"onfelder, (NA49 Coll.), Ph.D. thesis, TU M\"unchen, 1996.
\bibitem{W93}
  K. Werner, Phys. Rep. {\bf 232} (1993) 87.
\bibitem{SSG89}
  H. Sorge, H. St\"ocker, and W. Greiner, Ann. Phys. (N.Y.) {\bf 192}
  (1989) 266.
\bibitem{PSK92}
 Y. Pang, T.J. Schlagel, and S.H. Kahana, Phys. Rev. Lett. {\bf 68} 
 (1992) 2743; T.J. Schlagel, Y. Pang, and S.H. Kahana, Phys. Rev. Lett. 
 {\bf 69} (1992) 3290.
\bibitem{A97}
 J. Aichelin, Nucl. Phys. A (1997), in press (Los Alamos eprint 
 archive nucl-th/9609006).
\bibitem{MKF96}
  M. Martin, H. Kalechofsky, P. Foka, and U.A. Wiedemann, Los Alamos 
  eprint archive nucl-th/9612023.
\bibitem{S73}
 E. Shuryak, Phys. Lett. B{\bf 44} (1973) 387; Sov. J. Nucl. Phys.
 {\bf 18} (1974) 667.
\bibitem{YK78}
  F. Yano and S. Koonin, Phys. Lett. B{\bf 78} (1978) 556.
\bibitem{P94}
 S. Pratt et al., Nucl. Phys. A{\bf 566} (1994) 103c; and in 
 {\it Quark-Gluon Plasma 2}, ed. by R.C. Hwa (World Scientific, Singapore, 
 1995), p.~700.
\bibitem{CH94}
 S. Chapman and U. Heinz, Phys. Lett. B{\bf 340} (1994) 250.
\bibitem{PRW94}
 M. Pl\"umer, L.V. Razumov, and R.M. Weiner, Phys. Rev. D{\bf 49} 
 (1994) 4434; A. Timmermann, M. Pl\"umer, L.V. Razumov, and R.M. Weiner, 
 Phys. Rev. C{\bf 50} (1994) 3060.
\bibitem{PGG90}
 S. Padula, M. Gyulassy, and S. Gavin, Nucl. Phys. B{\bf 329} (1990) 357.
\bibitem{Z96}
 J. Zim\'anyi, talk given at the HBT96 Workshop, ECT* Trento, Sep. 16-27, 
 1996.
\bibitem{MP97}
 H. Merlitz and D. Pelte, Los Alamos eprint archive nucl-th/9702005.
\bibitem{GKW79}
 M. Gyulassy, S.K. Kauffmann, and L.W. Wilson, Phys. Rev. C{\bf 20}
 (1994) 2267.
\bibitem{Wetal}
 U.A. Wiedemann, P. Foka, H. Kalechofsky, M. Martin, C. Slotta, Q.H. Zhang,
 in preparation.
\bibitem{L89}
 B. L\"orstad, Int. J. Mod. Phys. A{\bf 12} (1989) 2861.
\bibitem{fn1}
 For a non-relativistic Boltzmann distribution, $\rho(x,p) \sim 
 \exp[-\bbox{p}^2/(2mT)]$, the prescription (\ref{14}) results in 
 an effective Boltzmann distribution with an increased temperature
 $T_{\rm eff} = T + 1/(2m\sigma^2)$ \protect\cite{Z96,MP97}.
\bibitem{GGLP60}
 G. Goldhaber, S. Goldhaber, W. Lee, and A. Pais, Phys. Rev. {\bf 120} 
 (1960) 300.

\end{thebibliography}
\end{document}